# LINAC*LHC BASED ep, γp, eA, γA and FELγ-A COLLIDERS

**Ö. Yavaş, A. K. Çiftçi**
Univ. of Ankara, Faculty of Science, Dept. of Physics, 06100 Tandogan, Ankara, TURKEY
**S. Sultansoy**
DESY, Notke Str. 85, D-22607 Hamburg, GERMANY
Univ. of Ankara, Faculty of Science, Dept. of Physics, 06100 Tandogan, Ankara, TURKEY
Institute of Physics, Academy of Sciences, H. Cavid Ave. 33, Baku, AZERBAIJAN

*Abstract*

Main parameters of different colliders which can be realized if a special 1 TeV energy linear electron accelerator or corresponding linear collider is constructed tangential to LHC are estimated. It is shown that $L_{ep}=10^{32}$cm$^{-2}$s$^{-1}$ at $\sqrt{s_{ep}}$=5.29 TeV can be achieved within moderate upgrade of LHC parameters. Then, γp collider with the same luminosity and √s=4.82 TeV can be realized using Compton backscattering of laser beam off the electron beam. Concerning the nucleus beam, $L*A=10^{31}$cm$^{-2}$s$^{-1}$ can be achieved at least for light and medium nuclei both for eA and γA options. Finally, colliding of the FEL beam from an electron linac with nucleus beams from LHC will give a new opportunity to investigate nuclear spectroscopy and photo-nuclei reactions.

## 1 INTRODUCTION

The center of mass energies which will be achieved at different options of this machine are an order larger than those at HERA are and ~3 times larger than the energy region of TESLA⊗HERA, LEP⊗LHC and $\mu$-ring⊗TEVATRON (see the review [1]). In principle, luminosity values are ~7 times higher than those of corresponding options of the TESLA⊗HERA complex due to higher energy of protons. Following [1-4] below we consider electron linac with $P_e$=60 MW (Table 1) and upgraded proton beam from LHC (Table 2). The reasons for choosing superconducting linac, instead of conventional warm linacs (NLC, JLC) or CLIC, are listed in [2].

## 2 MAIN PARAMETERS OF ep COLLIDER

According to Tables 1 and 2, center of mass energy and luminosity for this option are √s=5.29 TeV and $L_{ep}=10^{32}$cm$^{-2}$s$^{-1}$, respectively, and an additional factor 3-4 can be provided by the "dynamic" focusing scheme [5]. Further increasing will require cooling at injector stages (work on the subject is under development [6]). This machine, which will extend both the $Q^2$ –range and *x*-range by more than two order of magnitude comparing to those explored by HERA, has a strong potential for both SM and BSM research.

## 3 MAIN PARAMETERS OF γp COLLIDER

The advantage in spectrum of the back-scattered photons and sufficiently high luminosity (for details see ref. [7,8]), $L_{\gamma p}>10^{32}$cm$^{-2}$s$^{-1}$ at *z*=0, will clearly manifest itself in a searching of different phenomena. The physics search potential of this option is reviewed in [9]. For example, thousands di-jets with $p_t$>500GeV and hundreds thousands single W bosons will be produced, hundred millions of $b^*b$ and $c^*c$ pairs will give opportunity to explore the region of extremely small $x_g$ etc.

In Fig. 1 the dependence of luminosity on the distance *z* between interaction point (IP) and conversion region (CR) is plotted (for corresponding formulae see [8]). In Fig. 2 we plot luminosity distribution as a function of γp invariant mass $W_{\gamma p}=2\sqrt{E_\gamma E_p}$ at *z*=5 m. In Fig. 3 this distribution is given for the choice of $\lambda_e$=0.8 and $\lambda_0$=-1 at three different values of the distance between IP and CR.

## 4 MAIN PARAMETERS OF eA COLLIDER

In the case of LHC nucleus beam IBS effects in main ring are not crucial because of large value of $\gamma_A$. The main principal limitation for heavy nuclei coming from beam-beam tune shift may be weakened using flat beams at collision point. Rough estimations show that $L_{eA}\cdot A>10^{31}$cm$^{-2}$s$^{-1}$ can be achieved at least for light and medium nuclei [1,

10]. By use of parameters of nucleus beams given in Table 3 one has $L_{eC} \cdot A = 10^{31} \text{cm}^{-2}\text{s}^{-1}$ and $L_{ePb} \cdot A = 1.2 \cdot 10^{30} \text{cm}^{-2}\text{s}^{-1}$, correspondingly.

## 5 MAIN PARAMETERS OF γA COLLIDER

Limitation on luminosity due to beam-beam tune shift is removed in the scheme with deflection of electron beam after conversion [8]. The dependence of luminosity on the distance between IP and CR for γC and γPb options are plotted in Figs. 4 and 5, respectively. As it is seen from the plots, $L_{\gamma C} \cdot A = 0.8 \cdot 10^{31} \text{cm}^{-2}\text{s}^{-1}$ and $L_{\gamma Pb} \cdot A = 10^{30} \text{cm}^{-2}\text{s}^{-1}$ at $z = 5$ m.

The physics search potential of this option, as well as that of previous three options, needs more investigations from both particle and nuclear physics viewpoints.

## 6 FELγ–A COLLIDER

The ultra-relativistic ions will see laser photons with energy $\omega_0$ as a beam of photons with energy $2\gamma_A \omega_0$, where $\gamma_A$ is the Lorentz factor of the ion beam. For LHC $\gamma_A = (Z/A)\gamma_p = 7446(Z/A)$, therefore, $0.1 \div 10$ keV photons, produced by the linac based FEL, correspond to $0.5 \div 50$ MeV in the nucleus rest frame. The huge number of expected events [11] and small energy spread of colliding beams will give opportunity to scan an interesting region with adjusting of FEL energy.

Table 1. Parameters of special electron linac

| | |
|---|---|
| Electron energy, GeV | 1000 |
| No of electrons per bunch, $10^{10}$ | 0.7 |
| Bunch length, mm | 1 |
| Bunch spacing, ns | 100 |
| No of bunches per pulse | 5000 |
| Pulse Length, μs | 1000 |
| Repetition rate, Hz | 10 |
| Beam power, MW | 56 |
| Normalised emittance, $10^{-6}$m | 10 |
| Beta function at IP, cm | 200 |
| $\sigma_{x,y}$ at IP, μm | 3.3 |
| Beta function at CR, cm | 2 |
| $\sigma_{x,y}$ at CR, μm | 0.33 |

Table 2. Upgraded parameters of LHC proton beam

| | |
|---|---|
| Proton energy, GeV | 7000 |
| No of protons per bunch, $10^{10}$ | 40 |
| Bunch spacing, ns | 100 |
| Normalised emittance, $10^{-6}$m | 0.8 |
| Bunch length, cm | 7.5 |
| Beta function at IP, cm | 10 |
| $\sigma_{x,y}$ at IP, μm | 3.3 |

Table 3. Parameters of C and Pb beams

| | C | Pb |
|---|---|---|
| Nucleus energy, TeV | 42 | 574 |
| Particles per bunch, $10^{10}$ | 1 | 0.01 |
| Normalised emittance, $10^{-6}$m | 1.25 | 1.4 |
| Bunch length, cm | 7.5 | 7.5 |
| Beta function at IP, cm | 10 | 10 |
| $\sigma_{x,y}$ at IP, μm | 5.8 | 6.9 |
| Bunch spacing, ns | 100 | 100 |

## 7 SUMMARY

The proposed complex, if realised, will open new horizons for both the particle and the nuclear physics. Therefore, it is necessary to continue the efforts on both machine and physics search potential aspects.

This work is supported by Turkish State Planning Organization under the Grant No DPT-97K-120420 and DESY.

## 8 REFERENCES

1. S. Sultansoy, DESY-99-159 (1999).
2. M. Tigner, B.H. Wiik and F. Willeke, Proceedings of the 1991 IEEE Particle Accelerators Conference (6-9 May 1991, San Francisco), vol. 5, p. 2910.
3. B.H. Wiik, Proc. of the Int. Europhysics Conf. on High Energy Physics /22-23 July, Marseille, France), p.739
4. Proceedings of the First. Int. Workshop on Linac-Ring Type ep and γp Colliders (9-11 April 1997, Ankara), published in Turkish J. Phys. 22 (1998) 525-775.
5. R. Brinkmann and M. Dohlus, DESY-M-95-11 (1995).
6. A.K. Ciftci, S. Sultansoy and O. Yavas, in preparation.
7. S.F. Sultanov, IC/89/409, Trieste (1989).
8. A.K. Ciftci et al., Nucl. Instrum. Meth. A365 (1995) 317.

9. S. Sultansoy, Turkish J. Phys. 22 (1998) 575.
10. R. Brinkmann et al., DESY preprint 97-239 (1997).
11. H. Aktas et al., Nucl. Instrum. Meth. A428 (1999) 271.

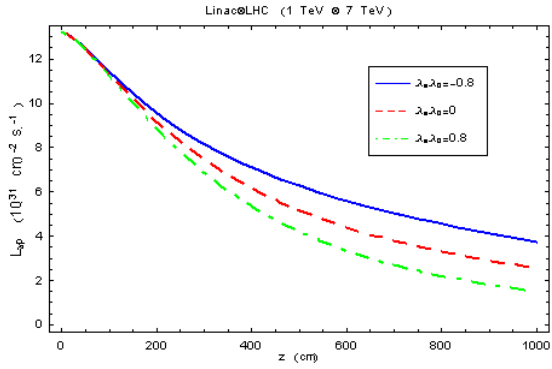

Figure 1

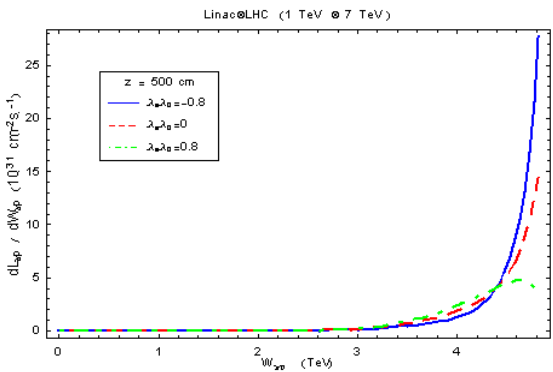

Figure 2

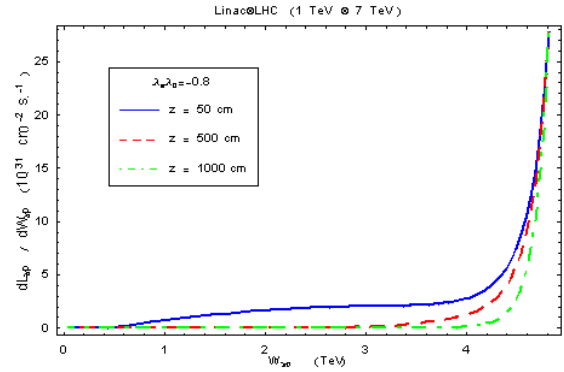

Figure 3

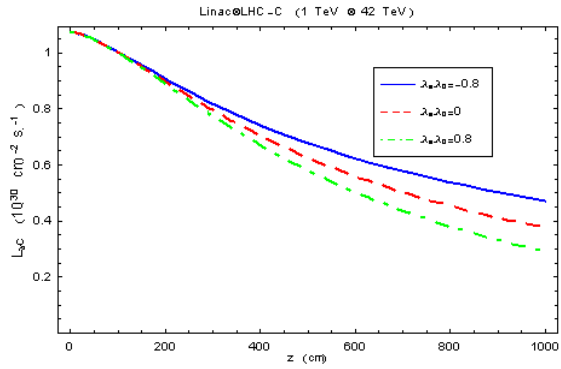

Figure 4

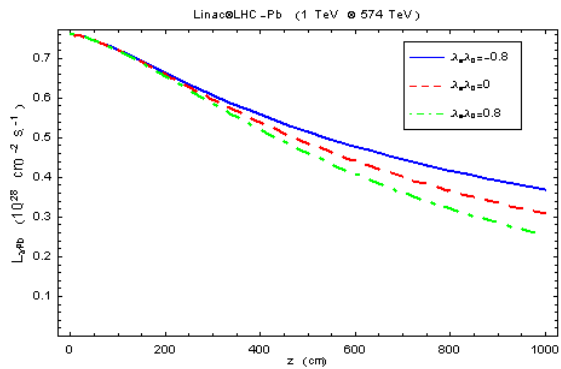

Figure 5